\documentclass[11pt,twoside]{article}

\usepackage{asp2006}
\usepackage{epsf}
\usepackage{psfig}
\usepackage{lscape}

\begin{document}
\title{Spectral Ages of Giant Radio Sources}   
\author{C. Konar,$^1$ D. J. Saikia,$^2$ M. Jamrozy,$^3$ and J. Machalski$^3$} 
\affil{$^1$IUCAA, Pune University Campus, Pune 411007, India \\
$^2$National Centre for Radio Astrophysics, TIFR, Pune 411007, India  \\  
$^3$Jagiellonian University, 30244 Krakow, Poland}

\begin{abstract} 
Multifrequency observations with the GMRT
and the VLA are used to determine
the spectral breaks in consecutive strips along the lobes of a sample of
selected giant radio sources (GRSs) in order to estimate their spectral ages.
The maximum spectral ages estimated for the detected radio emission in the lobes
of our sample of ten sources has a median value of $\sim$20 Myr.
The spectral ages of these GRSs are significantly older than smaller sources.
In all but one source (J1313+6937) the spectral age gradually increases
with distance from the hotspot regions, confirming that acceleration of the
particles mainly occurs in the hotspots. Most of the GRSs do not exhibit zero
spectral ages in the hotspots.  This is likely to be largely due to  contamination by more extended emission
due to relatively modest resolutions. The injection spectral indices range from $\sim$0.55 to
0.88 with a median value of $\sim$0.6.
We show that the injection spectral  index appears to be
correlated with luminosity and/or redshift as well as with linear size.
\end{abstract}


\section{Introduction}   
The radio continuum spectra in different parts
of an extended radio source contain important information about
the various energy losses and gains of the radiating
particles during the lifetime of the source. Giant radio sources (GRSs defined
to be larger than $\sim$1 Mpc, H$_o$=71 km s$^{-1}$ Mpc$^{-1}$, $\Omega_m$=0.27, $\Omega_{vac}$=0.73) are
suitable for classical spectral-ageing analysis due to their large angular extent
which can be covered by a significant number of resolution elements. 

However, while estimating spectral ages, caveats related to the evolution of
the local magnetic field in the lobes need to be borne in mind (e.g. Rudnick, Katz-Stone,
\& Anderson 1994; Jones, Ryu, \& Engel 1999; Blundell \& Rawlings 2000). Also, while Kaiser (2000)
has suggested that spectral and dynamical ages are comparable if
bulk backflow and both radiative and adiabatic losses are taken into account in a
self-consistent manner, Blundell \& Rawlings (2000) suggest that this may be so only
in the young sources with ages much less than 10 Myr. A comparison of the dynamical and
spectral ages show the dynamical ages to be
approximately a few times the maximum synchrotron ages of the emitting particles
(Machalski, Jamrozy, \& Saikia 2009 and references therein).

 \begin{figure}
   \vbox{
      \hbox{
    \psfig{file=J1313+696.610.ps,width=2.5in,angle=-90}
    \psfig{file=J1702+422.610.ps,width=2.5in,angle=-90}
            }
       \hbox{
    \psfig{file=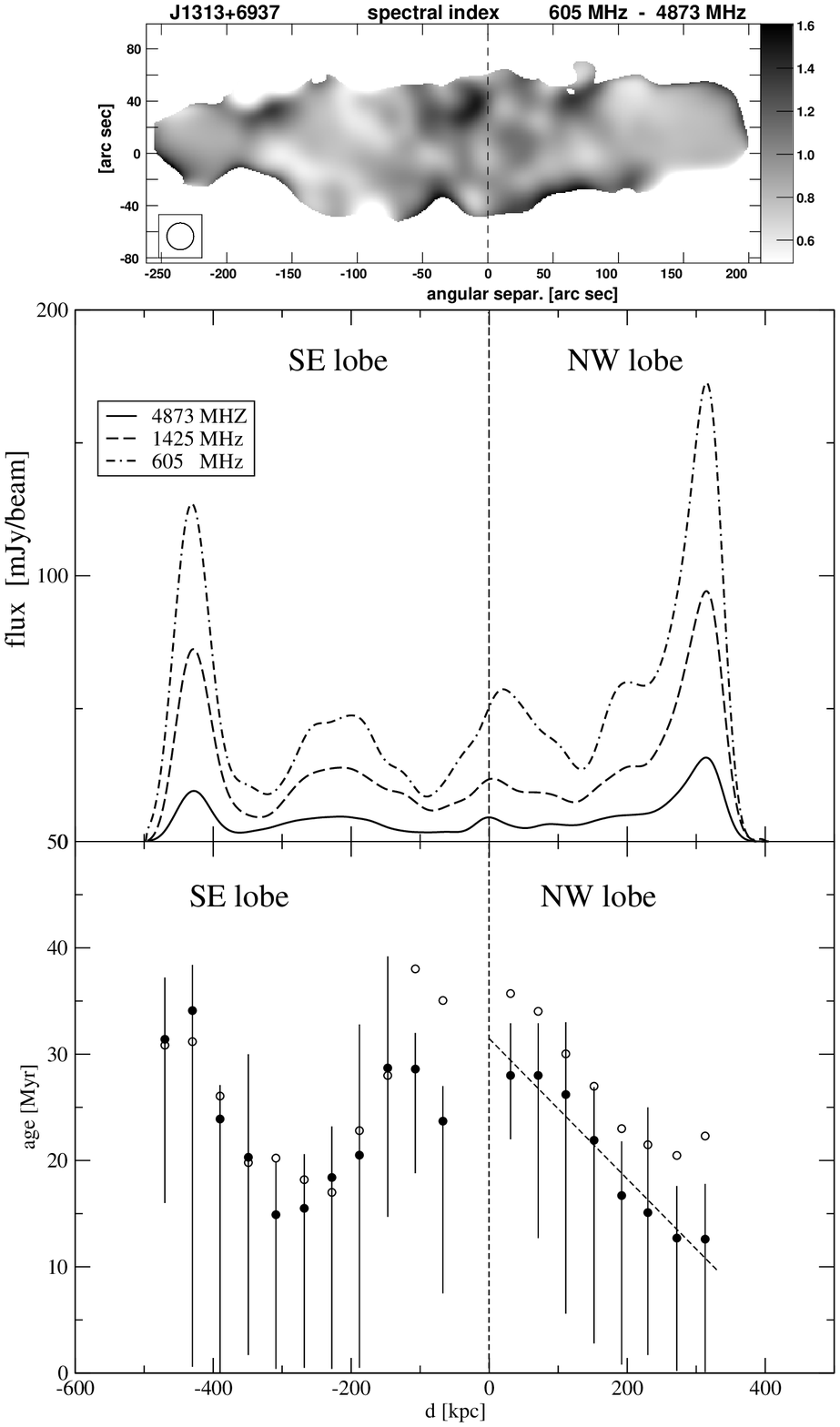,width=2.5in,angle=0}
    \psfig{file=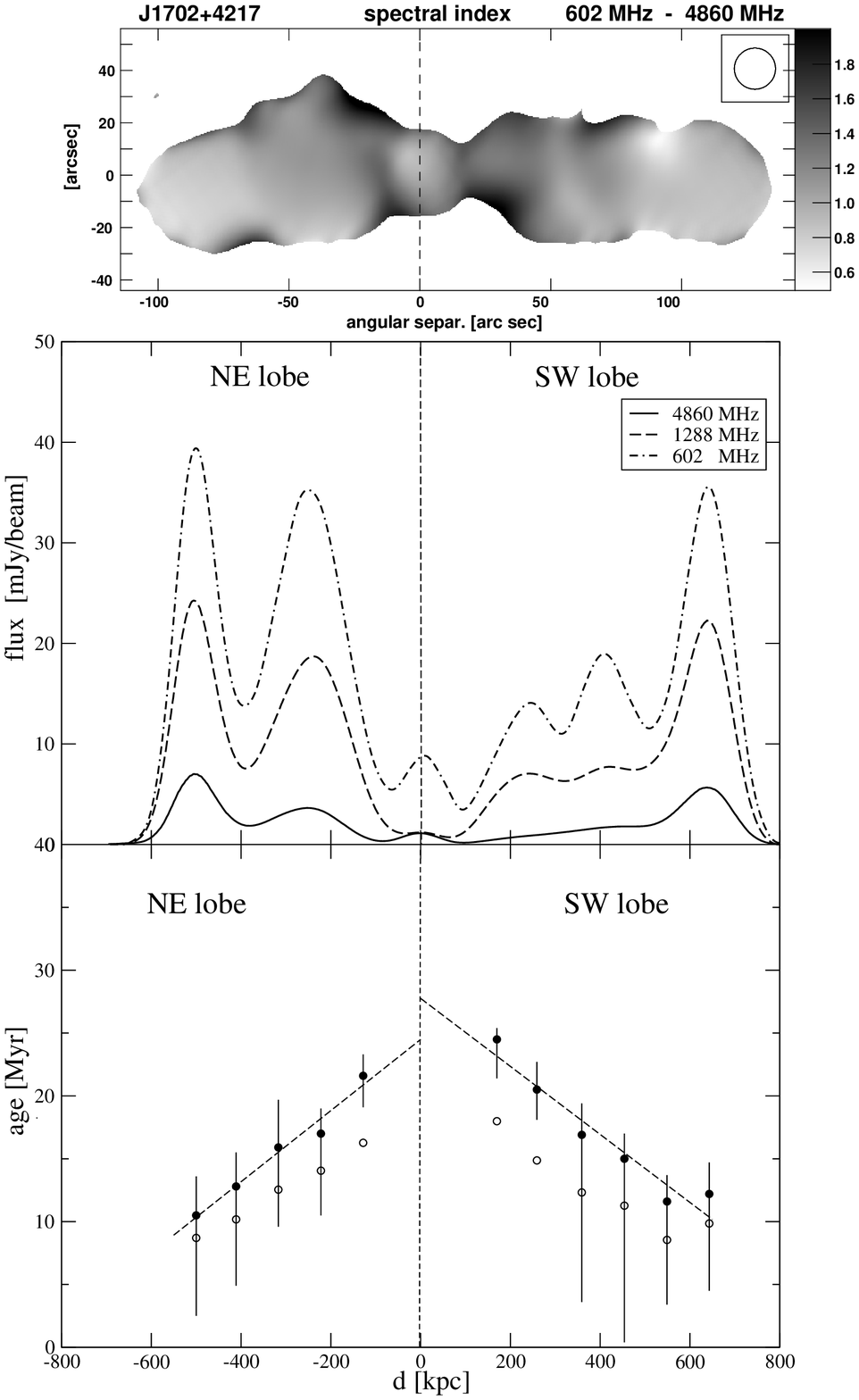,width=2.5in,angle=0}
        }
            }
 \caption{GMRT images of J1313+6937 and J1702+4217 at $\sim$600 MHz with an angular resolution of $\sim$6 and
7 arcsec respectively.
The $+$ sign indicates the position of the optical host galaxy of the source.
The ellipse at one of the corners indicates the resolution element. The middle and lower panels
show the spectral-index map, flux-density profiles, and the spectral age distributions.
The spectral ages have been estimated using the {\tt SYNAGE} package
(Murgia 1996; Murgia et al. 1999), and magnetic field values
determined using the Beck \& Krause (2005) formalism (filled circles and with error bars)
and the classical (e.g. Miley 1980) formalism (open circles without error bars).
For further details see Konar et al. (2008) and Jamrozy et al. (2008). }
 \end{figure}

We have made multifrequency observations with the GMRT and
the VLA, and estimated the spectral ages of ten GRSs (Konar et al. 2004, 2008; Jamrozy et al. 2008).
In this paper we summarise the results for two of these sources and discuss the dependence
of spectral age on size, and the relationships between the injection spectral index, $\alpha_{\rm inj}$
($S\propto\nu^{-\alpha}$), and luminosity and linear size.

\begin{figure}
\hspace{2.5cm}
\psfig{file=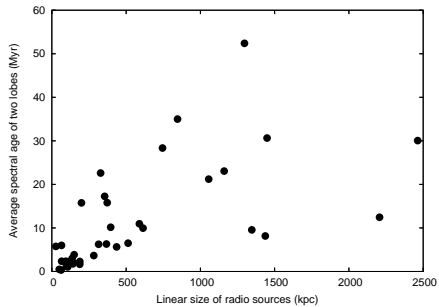,width=2.3in,angle=-90}
\caption{The spectral age as a function of the largest linear size.}
\end{figure}

\section{Spectral Ages of J1313+6937 and J1702+4217}
{\bf J1313+6937:} This source is at a redshift of 0.106 and has an overall linear size of 745 kpc,
somewhat smaller than 1 Mpc.
The GMRT image shows the prominent bridge of emission (Figure 1).
The spectral-index map and the intensity profiles
along the source axis suggest that the synchrotron radiation observed in separate
strips of this GRS may
be related to a mixture of emitting particles which were injected or accelerated
at different epochs. The SE lobe of this GRS is the only one, among the other
lobes studied in this piece of work in which the synchrotron ages determined do not more
or less gradually increase with distance from the hotspot area to the core.
Here $\alpha_{\rm inj}$ is 0.61 and the spectral age of both the lobes
using the classical equipartition magnetic field is $\sim$35 Myr.

\noindent
{\bf J1702+4217:}
The GRS J1702+4217 is at a redshift of 0.476 and has a largest projected size of 1160 kpc.
The GMRT 602-MHz image shows a prominent bridge with several peaks of emission (Figure 1).
Although there are peaks of emission visible in the lobe emission, especially at
the lower frequency, the spectral age increases smoothly with distance from the hotspots.
The spectral ages of the lobes, estimated for an injection spectral index of 0.59 are
$\sim$16 and 18 Myr for the north-eastern and south-western lobes respectively.

\subsection{Spectral Age$-$Linear Size Relation}
The maximum spectral ages estimated for the detected radio emission in the lobes
of our sources range from $\sim$6 to 46 Myr with a median of $\sim$23 Myr
using the classical equipartition fields. Using the magnetic field estimates from
the Beck \& Krause formalism the spectral ages range from $\sim$5 to 58 Myr with
a median of $\sim$24 Myr. 
The median linear size of these large sources is $\sim$1300 kpc. For 
a sample of extended 3CR sources with a median size of 342 kpc (Leahy, Muxlow, \& Stephens 1989), the spectral 
ages range from  $\sim$2.5 to 26 Myr with a median value of $\sim$8 Myr.
In the case of 14 compact 3CR sources with a median
size of 103 kpc (Liu, Pooley, \& Riley  1992) the values range from $\sim$0.3 to 5.3 Myr with a median value of
$\sim$1.7 Myr.  Considering these different samples, there is
a trend for the spectral ages to increase with linear size (Figure 2) as has been
noted earlier in the literature (e.g. Parma et al. 1999; Murgia et al. 1999).
The relative speeds of the lobe material 
range from $\sim$0.03 to 0.2c with a median of $\sim$0.1c for the
Liu et al. sample. The values for our GRSs are similar.

 \begin{figure}
   \hbox{
 \psfig{file=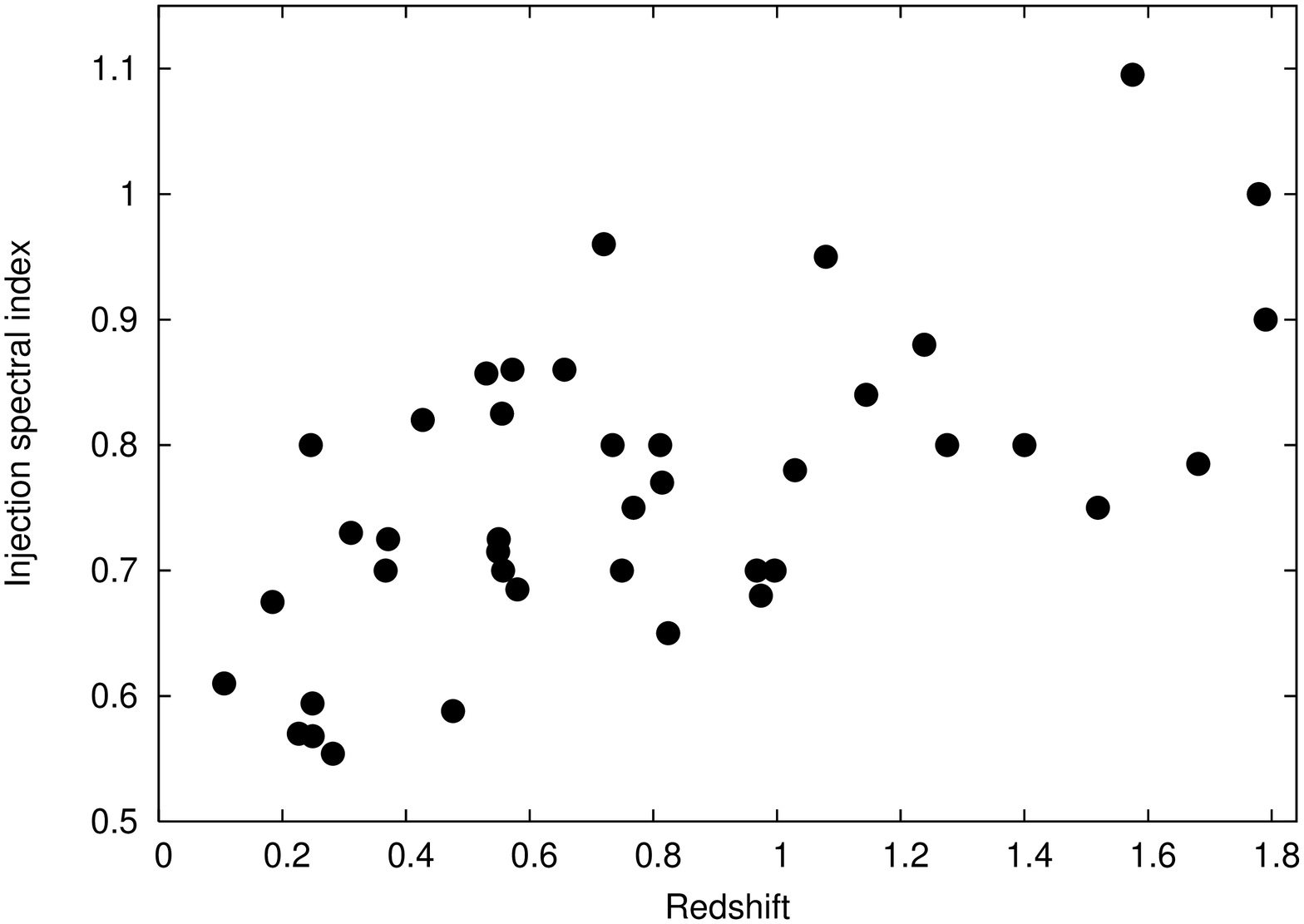,width=2.5in,angle=-90}
 \psfig{file=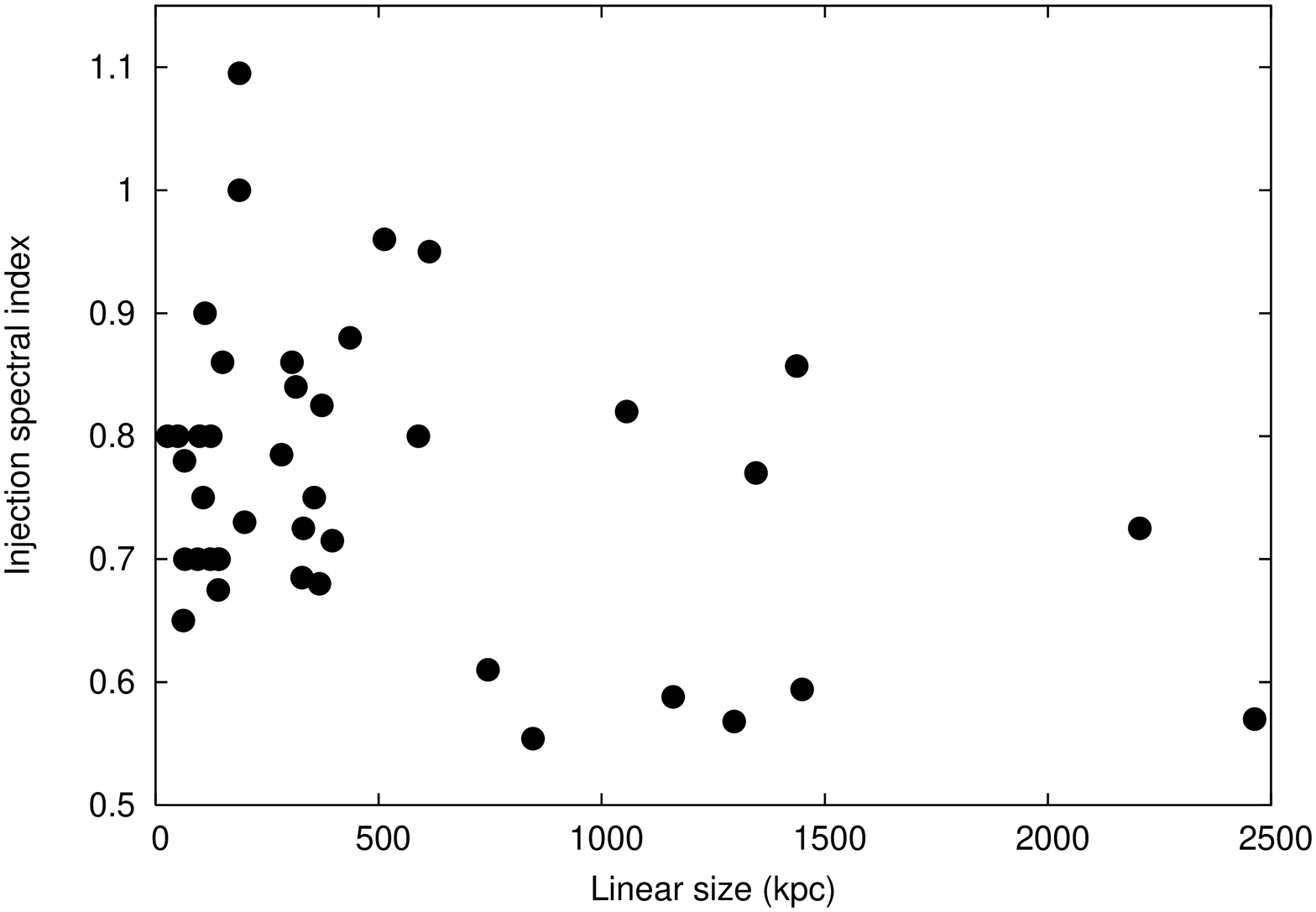,width=2.5in,angle=-90}
        }
 \caption{The correlation of the injection spectral index
$\alpha_{\rm inj}$ with redshift (left panel) and projected linear size
(right panel) for a sample of 3CR sources along with our sample of GRSs
(see Jamrozy et al. 2008).
           }
 \end{figure}

\section{Injection Spectral Indices}
The injection spectral indices range from 0.55 to 0.88 with a median value of
$\sim$0.6. Our estimates for the GRSs are marginally smaller than those estimated for
smaller sources by Leahy et al. (1989) and  Liu et al. (1992). Reliable
low-frequency measurements of the lobes using instruments such as GMRT, LWA and LOFAR are
required to get more reliable  estimates of the injection spectral indices. We have
explored possible correlations of $\alpha_{\rm inj}$ with other physical parameters
and find that it appears to increase with luminosity and/or redshift, but shows an
inverse correlation with linear size (Figure 3).



\end{document}